\documentclass[12pt]{iopart}
\usepackage{iopams}  

\usepackage{graphicx}% Include figure files
\usepackage{bm}% bold math
\usepackage[varg]{txfonts}

\begin{document}

\title{Anti-$E \times B$ flow field associated with a vortex formation
in a partially ionized plasma}

\author{Atsushi Okamoto \dag\footnote[1]{E-mail address: aokamoto@flanker.q.t.u-tokyo.ac.jp} , Kenichi Nagaoka \ddag ,
Shinji Yoshimura \ddag , Jovo Vranje\v{s} \S , Mitsuo Kono \$ ,
Shinichiro Kado \dag\ and Masayoshi Y. Tanaka \ddag
}

\address{\dag
High Temperature Plasma Center, The University of Tokyo,
2-11-16 Yayoi, Bunkyo, Tokyo 113-8656 Japan
}
\address{\ddag
National Institute for Fusion Science,
322-6 Oroshi, Toki, Gifu 509-5292 Japan
}
\address{\S
Center for Plasma Astrophysics,
Celestijnenlaan 200B, 3001 Leuven, Belgium
}
\address{\$
Faculty of Policy Studies, Chuo University,
Hachioji, Tokyo 192-0393, Japan
}

\begin{abstract}
 A high-density magnetized plasma has been studied for understanding of
 plasma dynamics in partially ionized plasmas.
 While research of plasma dynamics on fully ionized plasma has been
 developed in the past century as well as on weakly ionized plasma, the
 understanding of partially ionized plasma is a current issue especially
 in considering divertor plasmas, industrial plasmas and ionosphere
 plasma. 
 In this paper, the experimental result of ion flow field associated
 with a vortex formation in a partially ionized plasma is presented.
 The most remarkable result is that the direction of rotation is
 opposite to that of the ExB drift.

 The experiments have been performed in the high density plasma
 experiment (HYPER-I) device at the National Institute for Fusion
 Science.
 The HYPER-I device produces a cylindrical plasma (30 cm diameter and
 200 cm axial length) by electron cyclotron resonance (ECR) heating with
 a microwave (2.45 GHz, 80 kW maximum) launched at an open end of the
 chamber.
 The plasma density and the electron temperature are about
 $10^{13}\mathrm{cm}^{-3}$ and 5 eV, respectively, for the operation
 pressure 30 mTorr (Argon), which is several ten times higher than that
 of usual ECR plasma experiments.

 In the experimental condition described above, a tripolar vortex has
 spontaneously been formed.
 Ion flow velocity field, obtained with a directional Langmuir probe,
 shows that the rotational direction of each vortex is opposite to that
 of the $E \times B$ drift.
 Measurement of neutral density profile reveals that there is a steep
 density gradient of the neutrals around the vortex, thus inward
 momentum of the neutrals is generated due to the density gradient.
 The anti-$E \times B$ rotation is caused by the effective force
 attributable to radial momentum transfer from the directed neutrals to
 the ions with charge-exchange collision.
 The present experiment shows that this effective force may dominate the
 ambipolar-electric field and drive the anti-$E \times B$ vortical
 motion of ions.
\end{abstract}

\maketitle

\section{\label{sec:1}Introduction}
Understanding of plasma dynamics is an important and attractive issue in
plasma physics.
While research of plasma dynamics on fully ionized plasma has been
developed in the past century as well as on weakly ionized plasma,
recognition of common feature of dynamics observed in partially ionized
plasmas becomes a current topic.
Partially ionized plasma is defined as a plasma of intermediate
degree of ionization between fully ionized plasma and weakly ionized
plasma.
The ionization degree of partially ionized plasma is typically of the
order of ten percent, where adequate neutral particles and ions (and also
electrons) coexist and interact with each other.
Interesting phenomena related to plasma dynamics in partially ionized
plasmas have been observed; for example, flow reversal has been observed
in the scrape off layer of magnetically confined
plasmas,\cite{asakura2000prl}
ion upflow in ionospheric plasma.
Spatial dependence of the ionization degree\cite{vranjes2004pp} and the
role of neutral
particle itself are considered to be keys of these phenomena.
We have observed another phenomenon related to plasma dynamics in a
laboratory plasma with partially ionized condition.
The phenomenon is spontaneous formation of a tripolar vortex.
The tripolar vortex consists of an elliptic center vortex and two
bean-shaped satellites, which have opposite signs of polarity of
rotation to that of the center vortex, and remains stationary in time
during the whole discharge period.

Recently, tripolar vortices were observed in the ocean (the Bay of
Biscay) and in a rotating ordinary fluid to be self-organized from a
complex initial condition,\cite{pingree1992jgr} and from a forced
initial conditions.\cite{heijst1989nature, heijst1991jfm}
These results suggest that the tripolar vortex is a basic coherent structure
in rotating fluids\cite{kizner2004pre} or fluids subjected to the
Coriolis force.
A plasma in a magnetic field is equivalent to those fluids because the
Lorentz force has the same effect as the Coriolis force, and hence it
might be possible to occur a tripolar vortex in a plasma.\cite{vranjes2000pre}

In this paper, the experimental observation of a tripolar vortex in a
magnetized partially ionized plasma is presented.
A remarkable characteristic is that the tripolar vortex always appears
with a deep density depression of neutral particles and is confined in
its valley.
The flow velocity measurements revealed that each vortex rotates in the
anti-$E \times B$ direction, suggesting that there exists an effective
radial force acting on the ions, which overcomes the radial
electric field.
We propose that the charge-exchange collisions between the ions and
neutrals may produce the effective
force through the net momentum transfer.
When there is a strong inhomogeneity in the neutral density profile and the
charge-exchange collision is dominant, a directed momentum of the
flow of neutrals
is brought into the ions by the charge-exchange collision,
producing an effective force through the net momentum transfer.
It is shown that the effective force may dominate the ambipolar electric
field and drive the anti-$E \times B$ vortical motion of ions.
The tripolar vortex observed in the present experiments is considered to
be a neutral-induced tripolar vortex.
The existence of neutral particles usually causes a dissipative effect,
in which a dissipative instability\cite{chen} and a
modification of the mode
pattern take place.\cite{tanaka2001jpfr, kono2000prl}
It is worth pointing out that the existence of neutral particles may
change the ion dynamics qualitatively.
In Sec.~\ref{sec:2}, the experimental setup is described, and the
observation of tripolar vortex is presented in Sec.~\ref{sec:3}.
The mechanism of anti-$E \times B$ rotation is given in Sec.~\ref{sec:4}.

\section{\label{sec:2}Experimental Setup}
\begin{figure}[htbp]
\begin{center}
\includegraphics[width=156mm]{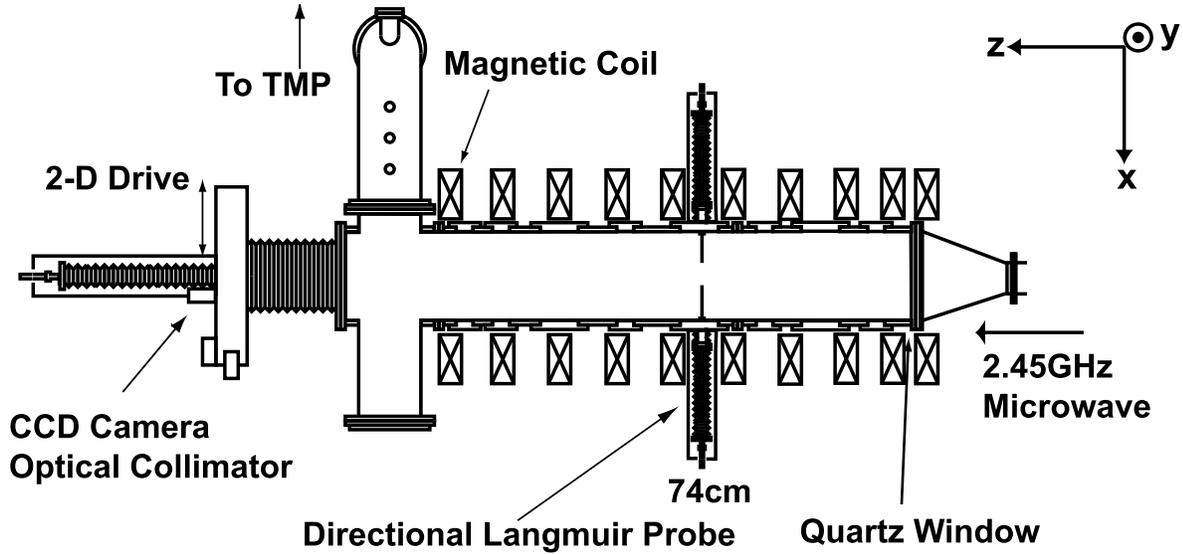}
\caption{\label{fig:hyper}A schematic of the HYPER-I device}
\end{center}
\end{figure}

The experiments have been performed in the high-density plasma experiment
(HYPER-I) device at National Institute for Fusion Science.\cite{tanaka1998rsi}
The HYPER-I device consists of a cylindrical chamber (30 cm in diameter,
200 cm in axial length) and ten magnetic coils, which produce magnetic
fields of $1-2$ kG along the chamber axis.
The schematic of HYPER-I device is shown in Fig.~\ref{fig:hyper}.
Plasmas are produced and sustained by the electron cyclotron resonance
(ECR) heating.
A microwave of frequency 2.45 GHz is generated by a klystron amplifier
(80 kW CW maximum) and is launched from an open end of the chamber, where
the high-field side condition
($\omega_\mathrm{ce}>\omega$, $\omega_\mathrm{ce}$: electron cyclotron
frequency, $\omega$: wave frequency)
is satisfied.
The magnetic field configuration is a so-called magnetic beach with
the ECR point at 90 cm from the microwave injection window.
An electron cyclotron wave is excited in the plasma and fully absorbed
before reaching the ECR point.\cite{tanaka1991jpsj}
The plasma dimension is 30 cm in diameter and 200 cm in axial length.
The typical electron densities are $10^{12} \, \mathrm{cm}^{-3}$ for the
operation pressure $1 \times 10^{-3} \, \mathrm{Torr}$ (Argon), and
$10^{13} \, \mathrm{cm}^{-3}$ for $3 \times 10^{-2} \, \mathrm{Torr}$.
The electron temperature gradually decreases with increasing the operation
pressure, and changes from 10 eV
($1 \times 10^{-3} \, \mathrm{Torr}$) to 3 eV
($3 \times 10^{-2} \, \mathrm{Torr}$).
The microwave input power in the present experiment is
$\leq 6.5 \, \mathrm{kW}$.

\begin{figure}[htbp]
\begin{center}
\includegraphics[width=75mm]{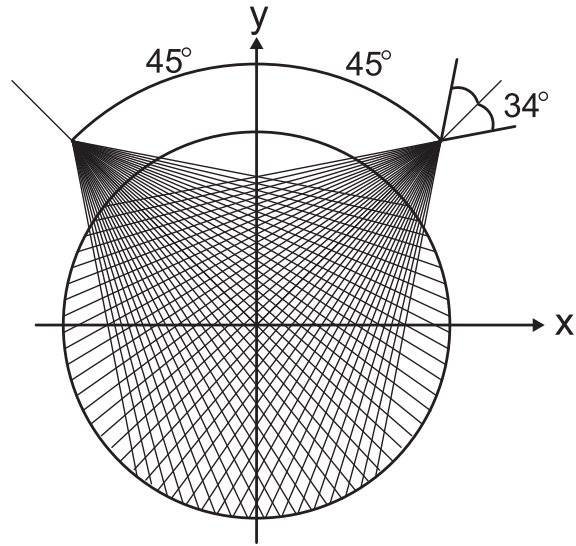}
\caption{\label{fig:2D_probe}Insertion chords of the two-dimensional DLP
 system. There are about 900 cross-points, on which the
 flow velocity vectors are determined.}
\end{center}
\end{figure}

The ion flow velocities have been measured with a directional Langmuir
probe (DLP), which collects a directed ion current through a small opening
(1 mm diam) made on the side wall of the ceramic insulator (3 mm diam).
The detailed structure of the DLP and its validity for measuring the ion
flow velocity is given in Ref.~\cite{nagaoka2001jpsj,okamoto2005jpfr}.
The flow velocity component at a certain angle $\theta$ with respect
to the reference axis, $v(\theta)$, is obtained by measuring two ion
saturation currents, $I_\mathrm{s}(\theta)$ and
$I_\mathrm{s}(\theta+\pi)$, and by using the following
relation:\cite{nagaoka2001jpsj}
\begin{equation}
  \frac{v(\theta)}{C_\mathrm{s}} =
  \frac{1}{\alpha}\frac{I_\mathrm{s}(\theta+\pi)-I_\mathrm{s}(\theta)}{I_\mathrm{s}(\theta+\pi)+I_\mathrm{s}(\theta)},
\end{equation}
where $\theta$ is the angle between the normal of electrode of the DLP and
the reference axis, and $C_\mathrm{s}$ is the ion sound speed.
$\alpha$ is a factor of the order of unity, and is calibrated by
cross-checking with other flow-measurement method such as
passive\cite{kado2004cpp} or laser-induced\cite{okamoto2004jpfr}
spectroscopy.
A two-dimensional velocity vector at a certain spatial point
$\bm{v}(\bm{r})$ on a plane perpendicular to the magnetic field is
determined by measuring the two components of the velocity vector.
A vector field plot of the flow velocity is constructed from a data set of
velocity components obtained by
a pair of DLPs mounted on the radial ports located at angles
$\pm 45^\circ$ from the vertical axis (see Fig.~\ref{fig:2D_probe}).
The insertion angle is changed up to $\pm 34^\circ$, and the insertion
chords with every 2 degrees of increment (decrement) are shown in
Fig.~\ref{fig:2D_probe}.
There are about $900$ cross-points in the cross section,
and the local velocity vectors are determined on each cross-point.
The distance between the nearest cross-points is about 7 mm
in the central region and about $3-11$ mm near the chamber wall.
From the original data set with the variable distances, we produce
a new data set on the lattice points with equal spacing
(5 mm) by interpolation, and then make the vector field plot of the flow
velocity.
The radial profile of density, electron temperature and space
potential have been measured with a Langmuir probe.

The spectroscopic measurements have been carried out to obtain the neutral
density profile.
A two-dimensional motor drive system with a collimated optic fiber is
equipped at the end of the chamber (see Fig.~\ref{fig:hyper}).
The focal point of the optical system is set to infinity to collect
the visible light emitted to the direction parallel to the axis of
the cylindrical plasma.
The diameter of the viewing chord is 6 mm, and the
collected light is analyzed by
a spectrometer (focal length 1 m) to select a
specific wavelength, and then detected by a photomultiplier tube.
We have observed the emission lines from the argon ions
[ $488.0\ \mathrm{nm}\ (4p\ ^2D^\circ-4s\ ^2P)$ ] and from the neutral argon
[ $425.9\ \mathrm{nm}\ (5p^\prime\ [\frac{1}{2}]-4s^\prime\ [\frac{1}{2}]^\circ)$ ].

The intensities of emitted spectral lines from the neutrals and ions are
respectively given by
$
I_\mathrm{Ar I}(r)  \propto \int n_\mathrm{n} n_\mathrm{e}
\langle\sigma^\mathrm{n}_\mathrm{ex} v\rangle \mathrm{d}l
$
and
$
I_\mathrm{Ar II}(r) \propto \int n_\mathrm{i} n_\mathrm{e}
\langle\sigma^\mathrm{i}_\mathrm{ex} v\rangle \mathrm{d}l
$,
where $\mathrm{d}l$ is the distance along the line of sight,
$\langle\sigma^j_\mathrm{ex} v\rangle (j=\mathrm{n,i})$ are the rate
coefficients of excitation process for the neutrals and ions, respectively.
When the radial density profiles of plasma and neutral particles are axially uniform,
the observed intensities $I_\mathrm{Ar I}(r)$ and $I_\mathrm{Ar II}(r)$
are proportional to the quantities
$n_\mathrm{n} n_\mathrm{e}$
and $n_\mathrm{i} n_\mathrm{e} \approx n_\mathrm{e}^2$, respectively.
Then the ratio of $I_\mathrm{Ar I}(r)$ to the square root of
$I_\mathrm{Ar II}(r)$ is proportional to the neutral density:
\begin{equation}
 \label{eq:opt_nn}
  \frac{I_\mathrm{Ar I}(r)}{\sqrt{I_\mathrm{Ar II}(r)}}
 \propto n_\mathrm{n}(r).
\end{equation}
We assume here the uniform profiles along the plasma axis,
which will be justified by the experimental results.

\section{\label{sec:3}Observation of Tripolar Vortex}
\begin{figure}[htbp]
\begin{center}
\includegraphics[width=156mm]{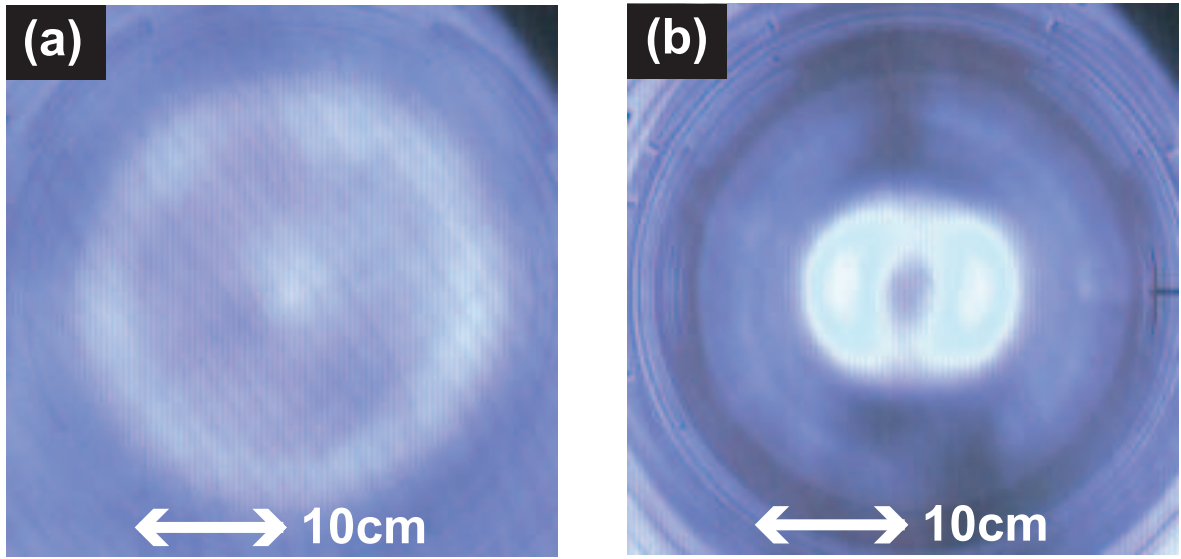}
\caption{\label{fig:image}End-view images of the plasma for different
 operation pressures.
 (a): $6.7\times 10^{-3} \, \mathrm{Torr}$,
 (b): $2.5\times 10^{-2} \, \mathrm{Torr}$.}
\end{center}
\end{figure}

End-view images of the plasma taken by a CCD camera for the different
operation pressures
are shown in Fig.~\ref{fig:image}, where Fig.~\ref{fig:image}(a) is for
$6.7 \times 10^{-3}  \, \mathrm{Torr}$ and Fig.~\ref{fig:image}(b) for
$2.5 \times 10^{-2}  \, \mathrm{Torr}$.
At the pressure $6.7 \times 10^{-3}  \, \mathrm{Torr}$ (and lower than this
pressure),
the plasmas are uniformly produced over the whole cross section of
the vacuum chamber.
At the operation pressure $2.5 \times 10^{-2}  \, \mathrm{Torr}$,
the two bright and bean-shaped regions appear in the central part of the plasma,
between which there is a dark elliptic region.
The diameter of the whole structure is about $15$ cm,
which is about one half of the diameter of the plasma filling in the
vacuum chamber.
This structure is spontaneously formed when
the microwave is turned on, and remains stationary in time during the whole
discharge period ($\sim 30 \, \mathrm{s}$).

\begin{figure}[htbp]
\begin{center}
\includegraphics[width=156mm]{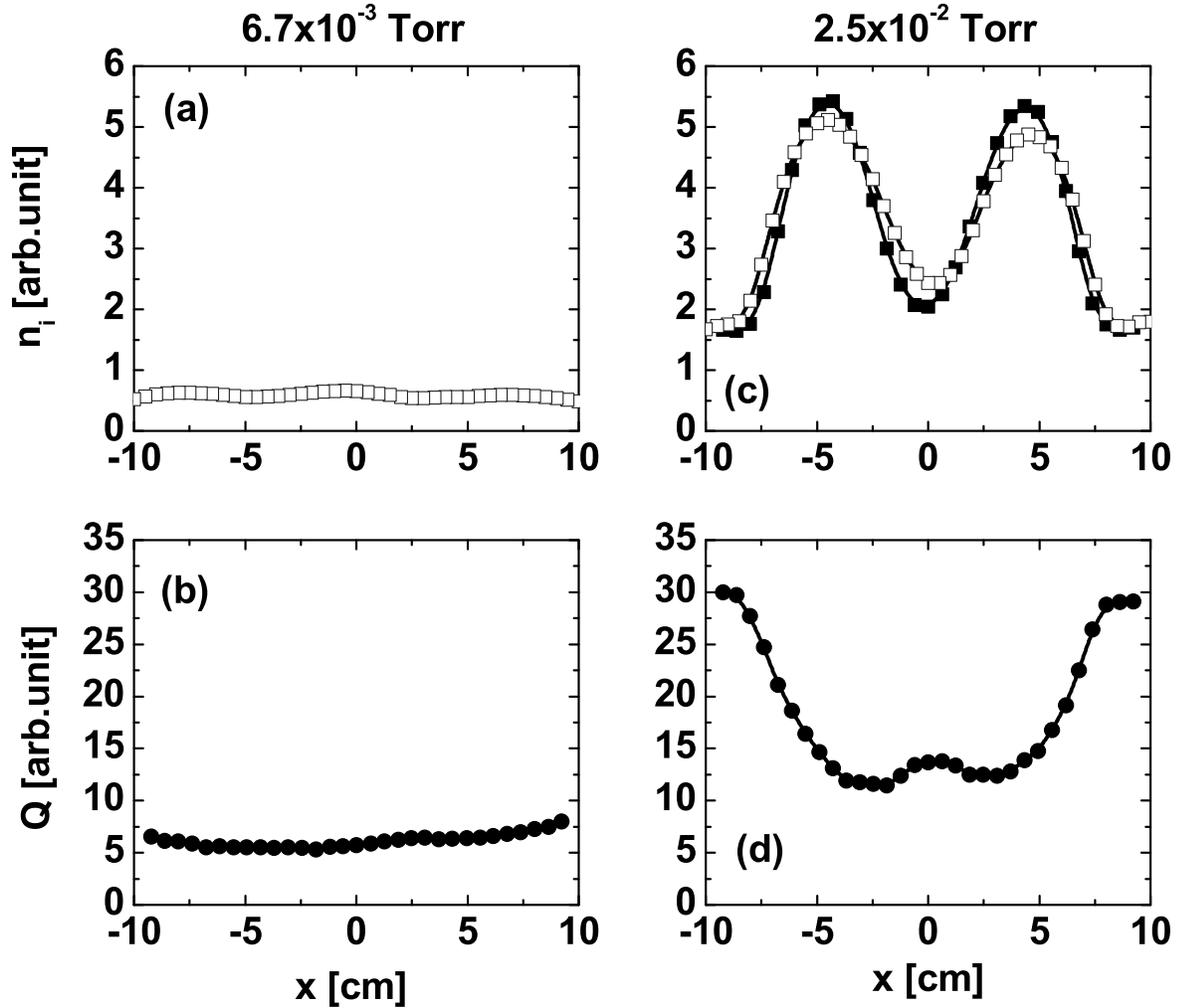}
\caption{\label{fig:density}Ion density profiles (top) and neutral
 density profiles (bottom). The operation pressures are the same as in
 Fig.~\ref{fig:image}.
 In Fig.~\ref{fig:density}(c), the ion density obtained from the
 optical measurement is indicated by closed square ($\blacksquare$), and
 the ion density measured with a Langmuir probe by open square
 ($\square$).
 In Figs.~\ref{fig:density}(b) and \ref{fig:density}(d), the quantity
 $Q$ is defined as
 $Q={I_\mathrm{Ar I}(r)}/{\sqrt{I_\mathrm{Ar II}(r)}}\propto n_\mathrm{n}$.}
\end{center}
\end{figure}

The radial profiles of the ion and neutral densities have been measured
along the horizontal center chord ($y=0$) for the same
operation pressures as in Fig.~\ref{fig:image}, and the results
are shown in Fig.~\ref{fig:density}.
The ion density profiles measured with a Langmuir
probe are indicated in the top
[Figs.~\ref{fig:density}(a) and \ref{fig:density}(c)] and the
corresponding neutral density profiles determined by
Eq.~(\ref{eq:opt_nn}) in the bottom
[Figs.~\ref{fig:density}(b) and \ref{fig:density}(d)].
The ion density profile determined by the optical measurement, which is
given by $n_\mathrm{i}(r) \propto \sqrt{I_\mathrm{Ar II}(r)}$, is also shown
in Fig.~\ref{fig:density}(c) to cross-check the validity of the optical
method.
The calibration factor for the optical measurement is determined so as
to give the equal value of $n_\mathrm{i}(r)$ at the point $r=10$ cm.
There is a good agreement between the two.
Since the optical measurement provides a line-integrated quantity, this
agreement suggests that the observed structure is axially homogeneous.
In fact, we have confirmed by the Langmuir probe measurement that the
observed structure is axially homogeneous at least more than 90 cm.
It has also been confirmed that the bean-shaped bright regions shown in
Fig.~\ref{fig:image}(b) are the ion density clumps.

In accordance with the appearance of bright spots, a neutral density
depression occurs in the central region of the plasma
[Fig.~\ref{fig:density}(d)].
It should be emphasized that when the
double-peaked structure is generated in the ion density profile, the neutral density
profile becomes a double-minimum distribution as shown in
Fig.~\ref{fig:density}(d).
Moreover, there is a close relation between the positions of the
steepest density gradient of the ions and the neutrals.
The width of ion clump and that of neutral depression
exhibit a quite well agreement,
suggesting that the observed ion density structure is
confined in the neutral density depression, and there might be a dynamical
coupling between the two.

\begin{figure}[htbp]
\begin{center}
\includegraphics[width=156mm]{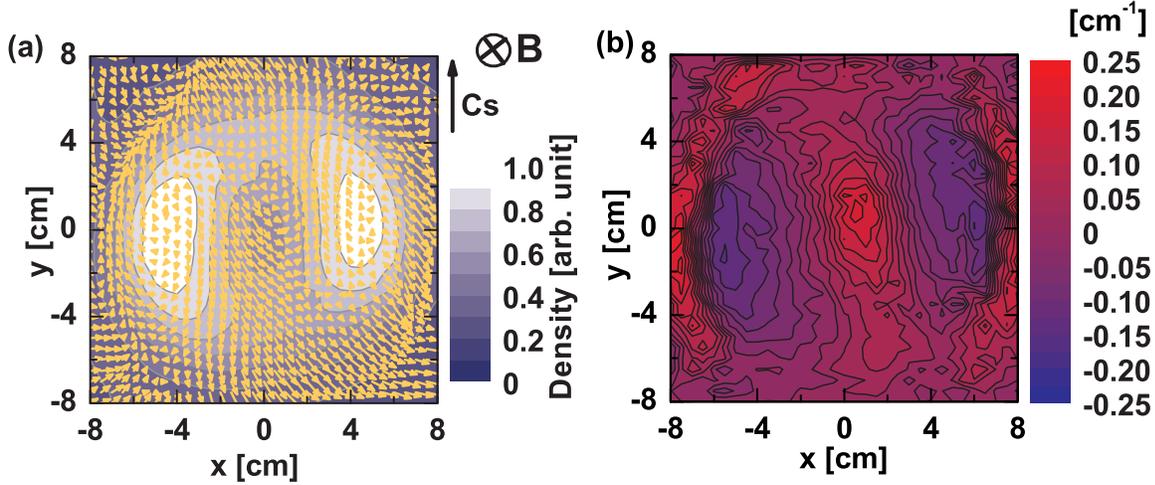}
\caption{\label{fig:tripole}
 (a): Vector field plot of the ion flow velocity and the density
 contour (background).
 The direction of the magnetic field is indicated in the upper right of
 the figure.
 The magnitude of flow velocities are normalized by the ion sound speed,
 which is also indicated in the upper right of the figure.
 (b): Contour plot of the $z$-component of
 vorticity.}
\end{center}
\end{figure}

The vector field plot of the ion flow velocity and the vorticity
distribution\cite{nagaoka2002prl} for a double-peaked structure are
shown in Fig.~\ref{fig:tripole}.
Figure~\ref{fig:tripole}(a) indicates the flow velocity field with the
contour map of ion density in the background, and Fig.~\ref{fig:tripole}(b) the
distribution of $z$-component of vorticity, which is constructed from
the velocity data by using the following equation:
$  \omega = (\mathrm{rot}\bm{v})_z
        \simeq {\oint\bm{v}_\perp\cdot\mathrm{d}\bm{l}}/{\Delta S} $.
In this calculation, the path of integration is so chosen to pass through
the four velocity vectors on the minimum lattice points of the velocity
field, and $\Delta S$ is the area surrounded by the integration path.
As seen in Fig.~\ref{fig:tripole}(a), there are two clockwise vortical
motions in both sides, which correspond to the ion density clumps.
Between these clumps, there presents a counterclockwise motion whose
center is a little bit shifted upward.
Since the velocity field pattern is a superposition of vortical motion
and other flows such as the diffusive flux, the vorticity
distribution contour is much more useful to recognize the existence of
vortices.\cite{okamoto2002jpfr}
Figure~\ref{fig:tripole}(b)
clearly shows the existence of three vortices; one vortex locates in
the center with a positive
polarity (counterclockwise rotation) and two satellites in the both
sides with negative polarities (clockwise rotation).
Therefore, the observed flow structure is a tripolar
vortex.\cite{okamoto2003pp}
Tripolar vortices have been already found in ordinary fluids.
Unlike these vortices in ordinary
fluids,\cite{heijst1989nature,heijst1991jfm} which slowly rotate as a
whole, the global vortex pattern in the plasma is stationary in the
laboratory frame, which suggests
that there exists a vortex solution with zero
eigenfrequency.
The existence of stationary tripolar solution in a plasma has been
shown in Ref.~\cite{vranjes2002prl}.

\section{\label{sec:4}Mechanism of Anti-$\bm{E} \bm{\times} \bm{B}$ rotation}
\begin{figure}[htbp]
\begin{center}
\includegraphics[width=85mm]{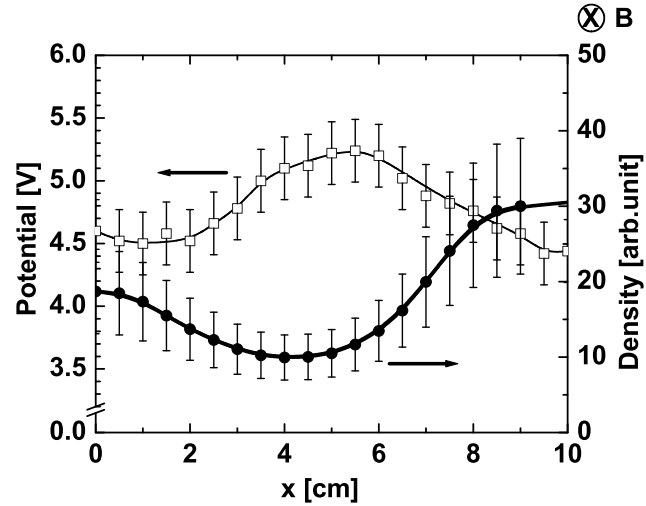}
\caption{\label{fig:potential}Electrostatic potential ($\square$) and
 neutral density profile ($\bullet$) of the tripolar vortex. The direction
 of the magnetic field is indicated in the
 upper right of the figure.}
\end{center}
\end{figure}

The electrostatic potential measurements revealed that the potential profile
along the horizontal chord ($y=0$)
is a double-peaked one, and the half of the profile is shown
in Fig.~\ref{fig:potential}, in which the corresponding neutral density
profile is also depicted.
In evaluating the neutral density $n_\mathrm{n}(r)$, the effect of
temperature variation along the horizontal chord
($\delta T_\mathrm{e}/T_\mathrm{e} \simeq 0.1$) on the atomic cross section
is taken into account.
As seen in the figure, the peak position of the potential ($x \simeq 5\mathrm{cm}$) coincides
with that of ion density, and its profile is convex around the density
peak, and concave near the center axis.
This means that the expected $E \times B$ rotations due to this potential
profile are counterclockwise for the satellite vortices,
and clockwise for the core vortex,
which are apparently opposite to the experimental observations.
There should be a radially inward force (or momentum transfer) to
explain the observed rotation of ions.
The probable mechanism of generation of inward force is a net momentum
transfer between the ions and neutrals.
As shown in Fig.~\ref{fig:density}, a steep density gradient in the neutrals
coexists with the tripolar vortex.
In this circumstance, there is a directed flow of neutrals induced by
the neutral density gradient, which may be given by
\begin{equation}
 \bm{v}_\mathrm{n} = -D_\mathrm{eff} \nabla\log n_\mathrm{n},
\end{equation}
where $n_\mathrm{n}$ is the density of neutrals.
Since the mean free path of neutrals is comparable to the scale of
vortex, we introduce an effective coefficient $D_\mathrm{eff}$
to include the enhancement factor.
When the charge-exchange interaction is dominant, the directed flow of
neutrals may become a source of inward momentum, i.e.,
during unit time interval, a momentum
$P_{\mathrm{n}\rightarrow\mathrm{i}}=\nu_\mathrm{ni}Mn_\mathrm{n}v_\mathrm{n}$
($\nu_\mathrm{ni}$:
charge-exchange collision frequency of neutrals with ions) is brought
into the ion fluids,
while in the same interaction the ions lose a momentum
$P_{\mathrm{i}\rightarrow\mathrm{n}}=\nu_\mathrm{in}Mn_\mathrm{i}v_\mathrm{i}$
($\nu_\mathrm{in}$:
charge-exchange collision frequency of ions with neutrals), resulting in
a net momentum transfer.
Under the present experimental conditions, the magnitude of
$P_{\mathrm{i}\rightarrow\mathrm{n}}$ and
$P_{\mathrm{n}\rightarrow\mathrm{i}}$ might be in the same order of magnitude.
The ion momentum equation is then written as\cite{vranjes2002prl}
\begin{equation}
 \label{eq:motion}
 M n_\mathrm{i}
  \left[ \frac{\partial \bm{v}_\mathrm{i}}{\partial t}
   + (\bm{v}_\mathrm{i}\cdot\nabla)\bm{v}_\mathrm{i} \right]
 =
  e n_\mathrm{i} (\bm{E}+\bm{v}_\mathrm{i}\times\bm{B})
 -\nabla p_\mathrm{i}
 -\nu_\mathrm{in} M n_\mathrm{i} \left( \bm{v}_\mathrm{i}
                                  + D_\mathrm{eff} \nabla\log n_\mathrm{n} \right),
\end{equation}
where $\nu_\mathrm{ni} n_\mathrm{n}= \nu_\mathrm{in} n_\mathrm{i}$ is used.
The perpendicular velocity component is given by
 \begin{eqnarray}
   \bm{v}_\perp = \frac{1}{\omega_\mathrm{ci}^2+\nu_\mathrm{in}^2}
   & \Bigl[ &
     \frac{e}{M}(
         \omega_\mathrm{ci} \bm{e}_z \times \nabla_\perp \phi
       - \nu_\mathrm{in} \nabla_\perp \phi
       )
     +v_{T_\mathrm{i}}^2 (
         \omega_\mathrm{ci} \bm{e}_z \times \nabla_\perp \log n_\mathrm{i}
       - \nu_\mathrm{in} \nabla_\perp \log n_\mathrm{i}
       ) \nonumber \\
&{}& +(
         \omega_\mathrm{ci} \nu_\mathrm{in} D_\mathrm{eff} \bm{e}_z \times \nabla_\perp \log n_\mathrm{n}
       - \nu_\mathrm{in}^2 D_\mathrm{eff} \nabla_\perp \log n_\mathrm{n}
       )
     \Bigr],
 \end{eqnarray}
where $\omega_\mathrm{ci}$ is the ion cyclotron frequency.
In deriving the above equation, the convective term
$\bm{v}\cdot\nabla \bm{v}$
is omitted for simplicity, which is justified when the ion flow velocity
is less than the ion sound speed.
On the horizontal chord ($y=0$),
$\partial/\partial r \gg (1/r)\partial/\partial\theta$
holds, and hence the azimuthal velocity at $y=0$ has the simplest form
to be compared with the observed velocity, which is written as
\begin{equation}
 \label{eq:vperp}
 \frac{v_y}{C_\mathrm{s}}
 = \frac{\omega_\mathrm{ci} C_\mathrm{s}}{
         \omega_\mathrm{ci}^2+\nu_\mathrm{in}^2}
   \left[
     \frac{\partial}{\partial r}
       \left( \frac{e\phi}{T_\mathrm{e}} \right)
    +\frac{T_\mathrm{i}}{T_\mathrm{e}}\frac{\partial}{\partial r}
      \left( \log n_\mathrm{i} \right)
    +\frac{D_\mathrm{eff}}{C_\mathrm{s}^2
                          / \nu_\mathrm{in} }\frac{\partial}{\partial r}
      \left( \log n_\mathrm{n} \right)
   \right].
\end{equation}
The first term represents the $E \times B$ drift, the second term the
diamagnetic drift, and the third term the $F \times B$ drift due to the
''effective pressure of neutrals''.
The diamagnetic drift velocity is small compared with the $E \times B$
drift velocity because $T_\mathrm{i}/T_\mathrm{e} \ll 1$.

\begin{figure}[htbp]
\begin{center}
\includegraphics[width=85mm]{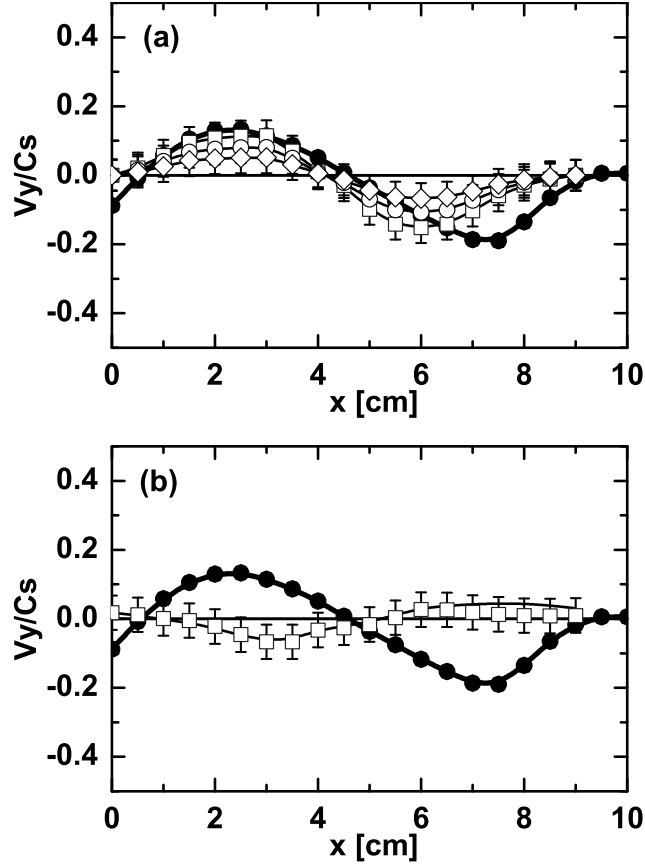}
\caption{\label{fig:velocity}Comparison between the azimuthal
 ion velocity measured with a DLP (bold solid line) and (a) the
 $F \times B$ drift velocity due to the neutral density gradient
 ($\square,\bigcirc,\diamondsuit$), and (b)
 the $E \times B$ drift velocity determined from the potential profile
 ($\square$).
 Note that the direction of $E \times B$ drift is opposite to the
 observed direction of rotation.
 In Fig.~\ref{fig:velocity}(a), the drift velocities for three different
 neutral temperatures are plotted
 ($\square : T_\mathrm{e}/T_\mathrm{n}=10 ,
 \bigcirc : T_\mathrm{e}/T_\mathrm{n}=20 ,
 \diamondsuit : T_\mathrm{e}/T_\mathrm{n}=50
 $).}
\end{center}
\end{figure}

Figure~\ref{fig:velocity} shows the azimuthal velocity profile along the
horizontal chord ($y=0$), where the bold solid line indicates the velocity
obtained with the DLP.
The $F \times B$ drift velocity due to the density gradient of neutrals,
which is determined by the third term of Eq.~(\ref{eq:vperp})
is also plotted in Fig.~\ref{fig:velocity}(a),
where $D_\mathrm{eff} = 5 D_\mathrm{c}$ ($D_\mathrm{c}$: collisional diffusion
coefficient) is used;
in evaluating the absolute value of neutral density at the tripolar
vortex region from the operation pressure measured at the chamber wall,
there is uncertainty of density depression.
This corresponds to an enhancement factor of several for $D_\mathrm{c}$
determined by the operation pressure.
In Fig.~\ref{fig:velocity}, we take $D_\mathrm{eff} = 5 D_\mathrm{c}$ as one
example. 
For comparison, the $E \times B$
drift velocity given by the first term of Eq.~(\ref{eq:vperp}) with the
observed potential profile is shown in Fig.~\ref{fig:velocity}(b).
In evaluating the $F \times B$ drift, we used the charge-exchange cross
section given by Chanin,\cite{chanin1957pr} and three different curves
corresponding $T_\mathrm{e}/T_\mathrm{n}=10,20,50$ are depicted.
Although there remains a certain ambiguity,
Fig.~\ref{fig:velocity}(a) shows a fairly good agreement between the
experimental observation and the $F \times B$ drift due to the density
gradient of neutrals.
It is emphasized that the $E \times B$ drift velocity determined by the
potential profile is opposite in direction [Fig.~\ref{fig:velocity}(b)].

\begin{figure}[htbp]
\begin{center}
\includegraphics[width=85mm]{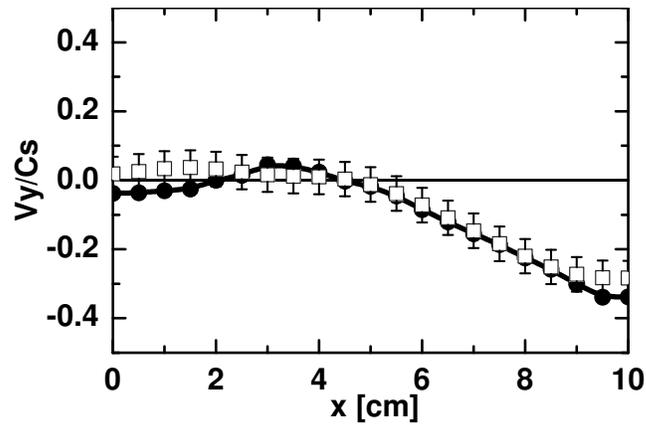}
\caption{\label{fig:lowp}
 Comparison between the azimuthal
 ion velocity measured with a DLP (bold solid line) and the $E \times B$
 drift velocity ($\square$) for the low pressure case
 ($6.7 \times 10^{-3}  \, \mathrm{Torr}$).}
\end{center}
\end{figure}

In the low pressure operations, however, the neutral particle
depression and thus the steep density gradient were not generated.
The observed azimuthal velocity in this case well agrees with that
determined by the $E \times B$ drift,\cite{nagaoka2001jpsj} which is
shown in Fig.~\ref{fig:lowp}.
It is interesting to note that the central region very weakly rotates
compared with the peripheral region, and there is a shear between the
two regions ($r=4-5$ cm), the position of which roughly coincides with
that of the center of
satellite vortex (see Fig.~\ref{fig:density} and~\ref{fig:tripole}).

We can conclude that the
anti-$E \times B$ rotation is attributable to the effective force
originated from the density gradient in neutrals.
The critical condition that the effective force overcomes the
electrostatic force may be given by
$
 e n_\mathrm{i} E \simeq
 \nu_\mathrm{in} M n_\mathrm{i} D_\mathrm{eff} \nabla \log n_\mathrm{n}
$.
Introducing $E\sim \phi/a$ ($a$: plasma radius) and
$\nabla\log n_\mathrm{n} \sim {l_\mathrm{cx}}^{-1}$ ($l_\mathrm{cx}$: mean free
path of charge-exchange collision), and using the appropriate values for
other quantities, we have
$
 l_\mathrm{cx}/a
 \sim
 (e\phi/T_\mathrm{e})^{-1}(\sigma_\mathrm{in}/\sigma_\mathrm{nn})\sqrt{T_\mathrm{n}/T_\mathrm{e}}
 \simeq 0.6 - 1.4
$,
where $D_\mathrm{eff} = 5 D_\mathrm{c}$ is assumed.
Experimentally, the pressure range of the transition to the
tripolar vortex is
$0.7 - 1.3 \times 10^{-2}  \, \mathrm{Torr}$,
which corresponds to $l_\mathrm{cx}/a \sim 2 - 3$,
showing a rough agreement with the above estimation.

% The mechanism of tripolar vortex formation may be considered as follows;
% an unperturbed plasma with velocity shear in the azimuthal component is
% subject to instability such as Kelvin-Helmholtz instability,
% and the most unstable mode grows and develops into a nonlinear
% structure.
% According to the theory,\cite{vranjes2002prl} the tripolar vortex is a
% coupled solution of unperturbed distribution ($m=0$) and $m=2$ mode.
% Probably, the $m=2$ mode is the most unstable mode in our case, as is
% often the case in many rotating
% fluids.\cite{heijst1989nature,heijst1991jfm,tanaka2001jpfr}
% Thus the developed $m=2$ mode superimposed on the unperturbed
% distribution produces the tripolar structure.

% As far as tripolar pattern is concerned, the existence of steep density
% gradient in neutrals doesn't affect the above scenario,
% and the occurrence of tripolar vortex is possible in
% $E \times B$-rotating plasmas as well.
% The present experimental condition may be considered to incidentally
% meet with a particular
% condition that a stationary tripolar vortex takes place in the laboratory
% frame.

\section{\label{sec:5}Conclusion}
A tripolar vortex has been observed in a magnetized plasma for the
first time.
The tripolar vortex coexists with a deep density hole of neutrals, and the
rotation direction is opposite to that of the $E \times B$ drift.
When a steep density gradient of neutrals is present, a net momentum transfer
may take place through the charge-exchange interaction, producing an
effective force acting on the ion fluids.
Our experiment shows that this effective force may overcome the radial
electric field, and generates an anti-$E \times B$ rotation.
Since the existence of neutrals usually brings a dissipation term
into the
equation of motion, it is of importance to note that the existence
of neutrals may qualitatively change the dynamical behavior
of ions.
The present result will be important in considering plasma behavior in
partially ionized plasmas such as ionospheric plasmas and surface
plasmas in confined systems.

\end{document}